\documentclass[aps,prl,reprint,showpacs,floatfix]{revtex4-1}

\usepackage{graphicx}
\usepackage{dcolumn}% Align table columns on decimal point
\usepackage{bm}% bold math

\usepackage{amssymb,amsfonts,amsmath}
\usepackage{chemarr}
\usepackage{color}
\usepackage{morefloats}

\usepackage{color} %  for colored comments

\newcommand{\app}{{\sc appendix }}
\newcommand{\ep}{\varepsilon}

\begin{document}

% Use the \preprint command to place your local institutional report
% number in the upper righthand corner of the title page in preprint mode.
% Multiple \preprint commands are allowed.
% Use the 'preprintnumbers' class option to override journal defaults
% to display numbers if necessary
%\preprint{}

%Title of paper
%\title{Analytical computation of frequency distributions of path-dependent processes by means of a non-multinomial maximum entropy approach}
\title{Fitting Power-laws in empirical data with estimators that work for all exponents}

\author{Rudolf Hanel$^1$}
%\email[]{rudolf.hanel@meduniwien.ac.at}
%\homepage[]{Your web page}
%\thanks{}
%\altaffiliation{}
%\affiliation{Section for Science of Complex Systems, Medical University of Vienna, Spitalgasse 23, 1090 Vienna, Austria}
\author{Bernat Corominas-Murtra$^1$}
%\email[]{bernat.corominas-murtra@meduniwien.ac.at}
%\homepage[]{Your web page}
%\thanks{}
%\altaffiliation{}
%\affiliation{Section for Science of Complex Systems, Medical University of Vienna, Spitalgasse 23, 1090 Vienna, Austria}
\author{Bo Liu$^1$}
%\email[]{bernat.corominas-murtra@meduniwien.ac.at}
%\homepage[]{Your web page}
%\thanks{}
%\altaffiliation{}
%\affiliation{Section for Science of Complex Systems, Medical University of Vienna, Spitalgasse 23, 1090 Vienna, Austria}
\author{Stefan Thurner$^{1,2,3,4}$}
\email{stefan.thurner@meduniwien.ac.at}
%\homepage[]{Your web page}
%\thanks{}
\affiliation{$^1$Section for Science of Complex Systems, Medical University of Vienna, Spitalgasse 23, 1090 Vienna, Austria}
\affiliation{$^2$Santa Fe Institute, 1399 Hyde Park Road, Santa Fe, NM 87501, USA}
\affiliation{$^3$IIASA, Schlossplatz 1, 2361 Laxenburg, Austria}
\affiliation{$^4$Complexity Science Hub Vienna, Josefst{\"a}dterstrasse 39, A-1090 Vienna, Austria} 
%Collaboration name if desired (requires use of superscriptaddress
%option in \documentclass). \noaffiliation is required (may also be
%used with the \author command).
%\collaboration can be followed by \email, \homepage, \thanks as well.
%\collaboration{}
%\noaffiliation

\date{\today}

\begin{abstract}
It has been repeatedly stated that maximum likelihood (ML) estimates of exponents of power-law distributions can only be reliably obtained for exponents smaller than minus one. The main argument that power laws are otherwise not normalizable, depends on the underlying sample space the data is drawn from, and is true only for sample spaces that are unbounded from above. Here we show that power-laws obtained from bounded sample spaces (as is the case for practically all data related problems) are always free of such limitations and maximum likelihood estimates can be obtained for arbitrary powers without restrictions. Here we first derive the appropriate ML estimator for arbitrary exponents of power-law distributions on bounded discrete sample spaces. We then show that an almost identical estimator also works perfectly for continuous data. We implemented this ML estimator and discuss its performance with previous attempts. We present a general recipe of how to use these estimators and present the associated computer codes. 
\end{abstract}

% insert suggested PACS numbers in braces on next line
%\pacs{02.70.Rr, 02.50.Cw, 05.10.Gg}
% ??? WHAT ARE THESE ???
% insert suggested keywords - APS authors don't need to do this
%\keywords{}

%\maketitle must follow title, authors, abstract, \pacs, and \keywords
\maketitle

% body of paper here - Use proper section commands
% References should be done using the \cite, \ref, and \label commands
%
%
%\subsection{Significance statement}
%
%$\\$$\\$
%
%{\bf Acknowledgements}.
%We greatly acknowledge the constructive comments of two anonymous referees who noted the relation to network diffusion and fragmentation. 
%This work has been supported by the Austrian Science Fund FWF under KPP23378FW.

%%%%%%%%%%%%%%%%%%%%%%%%%%%%%%%%%%%%%%%%%%%%%%%%%%%%%%%%%%%%%%%%%%%%%%%%%%%%%%%%%%%%%%%%
\section{Introduction}
The omnipresence of power-laws in natural, socio-economic, technical, and living systems has triggered immense research activity to understand their origins. It has become clear in the past decades that there exist several distinct ways to generate power-laws (or asymptotic power-laws), for an overview see for example \cite{newman,mitzenmacher}. In short, power-laws of the form
\begin{equation}
p(x) = C x ^{-\lambda} \quad, 
\label{pow}
\end{equation}
arise in critical phenomena \cite{physicsclassics,sornette}, in systems displaying self-organized criticality \cite{bak}, preferential attachment type of processes \cite{yuleproc,simon,prefattach,prefattach2}, multiplicative processes with constraints \cite{multproc}, systems described by generalized entropies \cite{tsallis,mgm14}, or sample space reducing processes \cite{BRSpnas2015}, i.e. processes that reduce the number of possible outcomes (sample space) as they unfold. Literally thousands of physical, natural, man-made, social, and cultural processes exhibit power-laws, the most famous being earthquake magnitudes \cite{earthquakes,earthquakes2}, city sizes \cite{citysizes,citysizes2}, foraging and distribution pattern of various animal species \cite{foraging}, evolutionary extinction events \cite{extinct}, or the frequency of word occurrences in languages, known as Zipf's law \cite{zipf}.

It is obvious that estimating power-law exponents from data is a task that sometimes should be done with high precision. For example if one wants to determine the universality class a given process belongs to, or when one estimates probabilities of extreme events. In such situations small errors in the estimation of exponents may lead to dramatically wrong predictions with potentially serious consequences. 

Estimating power-law exponents from data is not an entirely trivial task. Many reported power-laws are simply not exact power-laws, but follow other distribution functions. Despite the importance of developing adequate methods for distinguishing real power-laws from alternative hypotheses, we will not address this issue here since good standard literature on the topic of Bayesian {\em alternative hypotheses testing} exists, see for example \cite{press,jameso}. For power-laws some of these matters have been discussed also in \cite{plreview}. Here we simply focus on estimating power-law exponents from data on a sound probabilistic basis, using a classic Bayesian parameter estimation approach, see e.g. \cite{fisher}, that provides us with {\em maximum likelihood} (ML) estimators for estimating power-law exponents over the full range of reasonably accessible values. Having such estimators is of particular interest for a large classes of situations where exponents close to $\lambda\sim 1$ appear (Zipf's law). 
We will argue here that whenever dealing with data we can assume discrete and bounded samples spaces (domains), which guarantees that power-laws are normalizable for arbitrary powers $\lambda$. We then show that the corresponding ML estimator can then also be used to estimate exponents from data that is sampled from continuous sample spaces, or from sample spaces that are not bounded from above.

\subsection{Questions before fitting power-laws}

In physics the theoretical understanding of a process sometimes provides us with the luxury of knowing the exact form of the distribution function that one has to fit to the data. For instance think of critical phenomena such as Ising magnets in 2 dimensions at the critical temperature, where it is understood that the susceptibility follows a power-law of the form $(T-T_c)^{-\gamma}$, with $\gamma$ a critical exponent, that occasionally even can be predicted mathematically. However, often -- and especially when dealing with complex systems -- we do not enjoy this luxury and usually do not know the exact functions to fit to the data. 

In such a case, let us imagine that you have a data set and from first inspection you think that a power-law fit could be a reasonable thing to do. It is then essential, before starting with the fitting procedures, to clarify what one knows about the process that generated this data. The following questions may help to do so. 
\begin{figure}[t]
\centering
	\includegraphics[width=0.99\columnwidth]{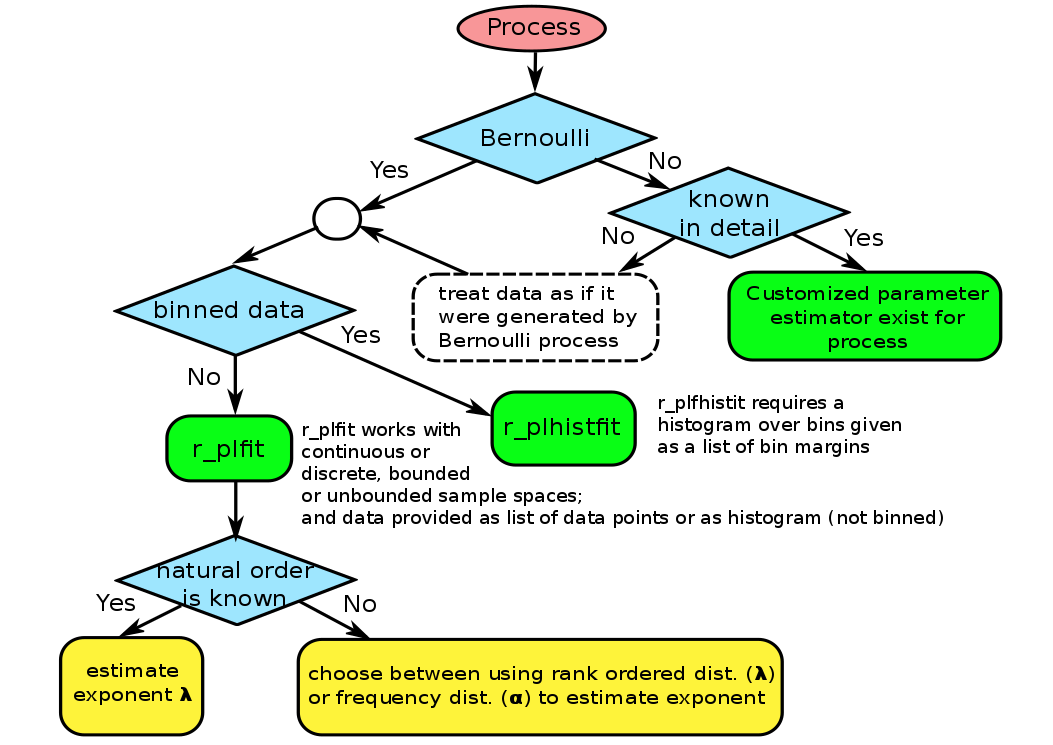}
\caption{Decision tree of questions that should be clarified before estimating power-law exponents from data. The tree shows under which conditions the fitting algorithms developed in this paper r$\_$plfit and r$\_$plhistfit can be used. 
 \label{fig:diagram}
 }
\end{figure}
\begin{itemize}
 \item Do you have information about the dynamics of the process that is generating what appears to be a power-law?
 \item Is the data generated by a Bernoulli process (e.g. tossing dice), or not (e.g. preferential attachment)? 
 \item Is the data available as a collection of samples (a list of measurements), or only coarse-grained in form of a histogram (binned or aggregated data). 
 \item Is the data sampled from a discrete (e.g. text) or continuous sample space (e.g. earthquakes)? 
 \item Does the data have a natural ordering (e.g. magnitudes of earthquakes), or not (e.g. word frequencies in texts)? 
\end{itemize}

The decisions one has to take before starting to estimate power-law exponents are shown as a decision-tree in  Fig. (\ref{fig:diagram}). If it is known that the process generating the data is not a Bernoulli process (for example if the process belongs to the family of history dependent processes such as e.g. preferential attachment), then one has the chance to use this information for deriving parameter estimators that are tailored exactly for the particular family of processes. If no such detailed information is available one can only treat the process as if it were a Bernoulli process, i.e. information about correlations between samples is ignored. If we know (or assume) that the data generation process is a Bernoulli process, the next thing to determine is whether the data is available as a collection of data points, or merely as coarse grained information in form of a histogram that collects distinct events into bins (e.g. histograms of logarithmically binned data).

If data is available in form of a data set of samples (not binned), a surprisingly general maximum likelihood (ML) estimator can be used to predict the exponent of an underlying power-law $p(x)\propto x^{-\lambda}$. This estimator that we refer to as ${\rm ML}^*$, will be derived in the main section. Its estimates for the underlying exponent $\lambda$, are denoted by $\lambda^*$. The code for the corresponding algorithm we refer to as \verb|r_plfit|. If information is available in form of a histogram of binned data, a different estimator becomes necessary. The corresponding algorithm (\verb|r_plhistfit|) is discussed in \app A and in the section below on discrete and continuous sample spaces. Both algorithms are available as matlab code \cite{rplfit}. 
For how to use these algorithms, see \app B.

If we have a dataset of samples (not binned), so that the \verb|r_plfit| algorithm can be used, it still has to be clarified whether the data has a natural order or not? Numerical observables such as earthquake magnitudes are {\em naturally} ordered. One earthquake is always stronger or smaller than the other. If observables are non-numeric, such as word types in a text, then a natural order can not be known {\em a priori}. The natural order can only be inferred approximately by using so-called {\em rank-ordering}; or alternatively -- by using the so-called {\em frequency distribution} of the data. Details are discussed below in the section on rank-order, frequency distributions, and natural order. 

Other issues to clarify are to see if a given sample space is continuous or discrete, and if the sample space is bounded or unbounded. These questions however, turn out to be not critical. One might immediately argue that for unbounded power-law distribution functions normalization becomes an issue for exponents $\lambda\leq 1$. However, this is only true for Bernoulli processes on {\em unbounded} sample spaces. Since all real-world data sets are collections of finite discrete values one never has to actually deal with normalization problems. Moreover, since most experiments are performed with apparati with finite resolution, most data can be treated as being sampled from a bounded, discrete sample space, or as binned data. For truly continuous processes the probability of two sampled values being identical is zero. Therefore, data sampled from continuous distributions can be recognized by sample values that are unique in a data set. See \app A for more details. 

Statistically sound ways to fit power-laws were  advocated and discussed in \cite{plreview,binneddata,acoral,Clcode}. They overcome intrinsic limitations of the  {\em least square} (LS) fits to logarithmically scaled data, which were and are widely (and often naively) used for estimating exponents. The ML estimator that was presented in \cite{plreview} we refer to as the ${\rm ML}_{\rm CSN}$ (for Clauset-Shalizi-Newman) estimator; its estimates for the exponent we denote by $\hat\lambda$. The approach that leads to ${\rm ML}_{\rm CSN}$ focuses on continuous data $x$ that follows a power-law distribution from Eq. (\ref{pow}), and that is bounded from below $x>x_{\rm min}>0$ but is not bounded from above (i.e. $x_{\rm max}>x$ with $x_{\rm max}=\infty$). In \cite{plreview} emphasis is put on how ML estimators can be used to infer whether an observed distribution function is likely to be a power-law or not. Also the pros and cons of using cumulative distribution functions for ML estimates are discussed, together with ways of treating discrete data as continuous data. For the continuous and unbounded case, simple explicit equations for the ${\rm ML}_{\rm CSN}$ estimator can be derived. The continuous approach however, even though it seemingly simplifies computations, introduces unnecessary self-imposed limitations with respect to the range of exponents that can be reliably estimated. ${\rm ML}_{\rm CSN}$ works brilliantly for a range of exponents between $-3.5$ and $-1.5$.

Here we show how to overcome these limitations -- and by doing so extend the accessible range of exponents -- by presenting the exact methodology for estimating $\lambda$ for discrete bounded data with the estimator ${\rm ML}^*$. While this approach appears to be more constrained than the continuous one we can show also theoretically that data from continuous and potentially unbounded sample spaces can be handled within essentially the same general ML framework as well. The key to the ${\rm ML}^*$ estimator is that it is not necessary to derive explicit equations for finding $\lambda^*$. Implicit equations in $\lambda$ exist for power-law probability distributions over discrete or continuous sample spaces that are both bounded from below {\em and} above. Solutions $\lambda^*$ can be easily obtained numerically. An implementation of the respective algorithms can be found in \cite{rplfit}, for a tutorial see \app B.

%%%%%%%%%%%%%%%%%%%%%%%%%%%%%%%%%%%%%%%%%%%%%%%%%%%%%%%%%%%%%%%%%%%%%%%%%%
\subsection{Rank-order, frequency distributions \& natural order}
\begin{figure}[t]
\centering
\includegraphics[width=0.95\columnwidth]{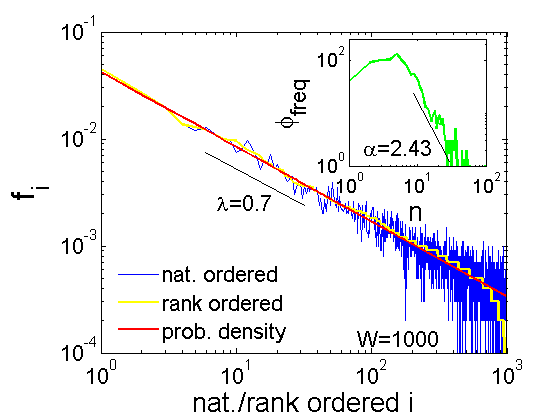}
\caption{The four types of distribution functions. Data is sampled from a power-law distribution $p(x)\propto x^{-\lambda}$ with an exponent $\lambda=0.7$ (red line). The relative frequencies $f_i$ are shown for $N=10000$ sampled data points according to their natural (prior) ordering that is associated with $p$ (blue). The rank-ordered distribution (posterior) is shown in yellow, where states $i$ are ordered according to their observed relative frequencies $f_i$. The rank-ordered distribution follows a power-law, except for the exponential decay that starts at rank$\sim 500$. A low frequency cut-off should be used to remove this part for estimating exponents. The inset shows the frequency distribution $\phi(n)$ that describes how many states $x$ appear $n$ times (green). The frequency distribution has a maximum and a power-law tail with exponent $\alpha=1+1/\lambda\sim2.43$. To estimate $\alpha$, one should only consider the tail of the frequency distribution function. 
\label{fig:rankvsfreq}
}
\end{figure}

%%%%%%%%%%%%%%%%%%%%%%%%%%%%%%%%%%%%%%%%%%%%%%%%%%%%%%%%%%%%%%%%%%%%%%%%%%
There exist three distinct types of distribution functions that are of interest in the context of estimating power-law exponents:  \\
\begin{description}
\item[i] The {\em probability distribution} $p(x)$ assigns a probability to every observable state-value $x$. Discrete and bounded sample spaces are characterized by $W$ state-types $i=1,\cdots,W$, with each type $i$ being associated with a distinct value $x=z_i$.
\item[ii] The {\em relative frequencies}, $f_i=k_i/N$, where $k_i$ is the number of times that state-type $i$ is observed in $N$ experiments. 
$k=(k_1,\cdots,k_W)$ is the {\em histogram} of the data. As explained below in detail, the relative frequencies can be ordered in two ways.\\
$\bullet$ If $f_i$ is ordered according to their descending magnitude this is called the {\em rank ordered} distribution.\\
		
$\bullet$ If $f_i$ is ordered according to the descending magnitude of the probability distribution $p(z_i)$, 
then they are {\em naturally ordered} relative frequencies.
\item[iii] The {\em frequency distribution} $\phi(n)$ counts how many state-types $i$ fulfill the condition  $k_i=n$.
\end{description}
In Fig. (\ref{fig:rankvsfreq}) we show these distribution functions. There $N=10000$ data points are sampled from $x\in\{1,\cdots,1000\}$, with probabilities $p(x)\propto x^{-0.7}$. The probability distribution is shown (red).  The relative frequency distribution $f$ is plotted in natural order (blue), the rank-ordered distribution is shown with the yellow line, which clearly exhibits an exponential decay towards the the tail. The inset shows the frequency distribution $\phi(n)$ of the same data. 
We next discuss how different sampling processes can be characterized in terms of natural order, rank-order, or frequency distributions. 

\subsubsection{Processes with naturally ordered observables}
For some sampling processes the ordering of the observed states is known. For example think of $x$ representing the numerical values of earthquake magnitudes. Here any two observations $x$ and $x'$ can be ordered with respect to their numerical value, or their {\em natural order}. Since power-law distributions  $p(x)\propto x^{-\lambda}$ are monotonic this is equivalent to ranking observations according to the probability distribution $p$ they are sampled from: 
The most likely event has {\em natural rank} $1$, the second most likely rank $2$, etc. In other words, we can order state-types $x$ in a way that over the sample space $\Omega=\{ z_i|i=1,\cdots,W\}$, $p=(p(z_1),\cdots,p(z_W))$ is a monotonic and decreasing function.

\subsubsection{Processes with rank-ordered observables}
If $p$ is not known {\em a priori} because the state-types $i$ have no numerical values $z_i$ attached, as happens for example with words in a text, we can only count relative frequencies $f_i$ (a normalized histogram) of states of type $i$, {\em a posteriori}, i.e. after sampling. To be clear, let $k=(k_1,\cdots,k_W)$ be the histogram of $N$ recorded states. $k_i$ is the number of times we observed type $i$, then $f_i=k_i/N$ is the relative frequency of observing states of type $i$. After all samples are taken, one can now order states with respect to $f_i$, such that the rank $1$ is assigned to state $i$ with the largest $f_i$, rank $2$ to $i'$ with the second largest $f_{i'}$, etc. $f=(f_1,\cdots,f_W)$ is called the {\em rank-ordered} distribution of the data.

The natural order imposed by $p$ and the rank-order imposed by $f$ are not identical for finite $N$. However, if data points have been sampled independently, then $f$ converges toward $p$ (for $N\to\infty$) and the rank-order induced by $f$ will asymptotically approach the natural order induced by $p$. The highest uncertainty on estimating the order induced by $p$ using $f$ is associated with the least frequent observations. Therefore, when estimating exponents from rank-ordered distributions, one might consider to use a low-frequency cut-off to exclude infrequent data.  

\subsubsection{Frequency distributions}
Exponents of power-laws can also be estimated from {\em frequency} distributions $\phi(n)$. These counts how many distinct state-types $i$ occur exactly $n$ times in the data. It does not depend on the natural (prior) order of states and therefore is sometimes preferred to the (posterior) rank-ordered distribution. However, complications may appear also when using $\phi(n)$. The frequency distribution $\phi(n)$ that is associated with a power-like 
probability distribution $p \propto x^{-\lambda}$ (and asymptotically to $f$) is not an exact power-law but a non-monotonic distribution (with a maximum). Only its tail decays as a power-law, 
$\phi(n)\propto n^{-\alpha}$. The exponents $\lambda$ and $\alpha$ are related through the well known equation
\begin{equation}
	\alpha=1+1/\lambda \quad .
	\label{alphalambda}
\end{equation}
If the probability distribution has exponent $\lambda$, the tail of the associated frequency distribution has exponent $\alpha$. Since the frequency distribution behaves like a power-law only in its tail, estimating $\alpha$ makes it necessary to constrain the observed data to large values of $n$. Note that this is equivalent to using a low-frequency cut-off. One option to do that is to derive a maximum entropy functional for $\phi(n)$ and fit the resulting (approximate) max-ent solution to the data. We do not follow this route here.

If the natural order of the data is known, one can directly use the natural ordered data in the ML estimates for the exponents. If it is not known, either the rank-ordered distribution can be used to estimate $\lambda$, or the frequency distribution to estimate $\alpha$, see Fig. (\ref{fig:diagram}).

One might also estimate both, $\lambda$ in the rank ordered distribution, and $\alpha$ in the frequency distribution of the data. Using Eq. (\ref{alphalambda}) to compare the two estimates may be used as a rough quality-check. If estimates do not reasonably coincide one should check whether the used data ranges have been appropriately chosen. If large discrepancies remain between $\alpha$ and 
$1+1/\lambda$ this might indicate that the observed distribution function in question is only an approximate power-law, for which Eq. (\ref{alphalambda}) need not hold. For a tutorial on how to use \verb|r_plfit| to perform estimates see \app B.  

%%%%%%%%%%%%%%%%%%%%%%%%%%%%%%%%%%%%%%%%%%%%%%%%%%%%%%%%%%%%%%%%%%%%%%%%%%
\subsection{Discrete and continuous sample spaces \& normalization}
%%%%%%%%%%%%%%%%%%%%%%%%%%%%%%%%%%%%%%%%%%%%%%%%%%%%%%%%%%%%%%%%%%%%%%%%%%
Data can originate from continuous sample spaces $\Omega_c=[x_{\rm min},x_{\rm max}]$, or discrete ones $\Omega_d=\{z_1,z_2,\cdots,z_W\}$. To each state-type $i=1,\cdots,W$, there is assigned a state-value $z_i$. Whether a distribution function $p(x)=  Z_\lambda^{-1} x^{-\lambda}$, with $x\in\Omega$, is normalizable or not, can only be decided once the %{\re nature of the} 
sample space $\Omega$ has been specified. The normalization factors for continuous and discrete $\Omega$ are 
\begin{equation}
\begin{array}{lclcl}
	Z_\lambda(\Omega_c)&\equiv&\int_{x_{\min}}^{x_{\max}} dx\ x^{-			
    \lambda}&=&\frac{x_{\max}^{1-\lambda}-x_{\min}^{1-\lambda}}{1-\lambda}
	\\&&\\
	Z_\lambda(\Omega_d)&\equiv&\sum_{x\in\Omega_d} x^{-\lambda}&=&\sum_{i=1}^{W} 	 z_i^{-\lambda}\quad .
\end{array}
\label{norm}
\end{equation}
For bounded sample spaces with  $0<x_{\min}\leq x\leq x_{\max}<\infty$, power-laws are always normalizable for arbitrary exponents $\lambda$, and a well defined ML estimator of $\lambda^*$ exists (see below). The normalization constants in Eq. (\ref{norm}) can be specified in \verb|r_plfit| (see \app B). 

Data sampled from a continuous sample space $\Omega_c$ can essentially be treated as if it were sampled from a discrete sample space $\Omega_d$, where $x\in\Omega_d$ are given by the unique collection of distinct values in the data set. That is, the data set $x=(x_1,\cdots,x_N)$ contains $N$ data points $x_n$ (that have $W$ unique values $z_i$, the states of type $i$) which we collect in the discrete sample space $\Omega=\{z_1,\cdots,z_W\}$. For truly continuous data we have $N=W$, since the probability of $x_m=x_n$ for $n\neq m$ is vanishing. As a consequence the histogram $k_i$, which counts the number of times $z_i$ appears in the data, 
%will essentially look like 
is essentially given by
$k_i=1$ for all $i=1,\cdots,W$. This provides us with a practical criterion for when to use the normalization constant for discrete or continuous data. For details see \app A.   

The equation for the ML estimator ${\rm ML}^*$, that yields the estimate $\lambda^*$, only requires the knowledge of the relative frequency distribution $f_i=k_i/N$ (in natural- or rank-order) of the observed state-types $i$, as we will see in Eq. (\ref{bayes4}) below. Therefore \verb|r_plfit| can work either with data sets $x$ or histograms $k$ over the unique values in the data sets. If data comes in coarse grained form, i.e. histograms, where each bin may contain a whole range of observable values $x$, then an estimator is required that is different from ${\rm ML}^*$  \cite{binneddata}, see also \app A. The corresponding code \verb|r_plhistfit| can also be downloaded from \cite{rplfit}. 

%%%%%%%%%%%%%%%%%%%%%%%%%%%%%%%%%%
\section{The ${\rm ML^*}$-estimator for power-laws from discrete sample spaces}
%%%%%%%%%%%%%%%%%%%%%%%%%%%%%%%%%%
Consider a family of random processes $Y$ that is characterized by the parameters $\theta=(\theta_1,\cdots,\theta_R)$. Let $Y$ be defined on a discrete sample space $\Omega=\{z_1,z_2,\cdots,z_W\}$, with $0<z_i<\infty$. The process $Y$ samples values $x\in \Omega$ with probability, 
\begin{equation}
	p(x|\theta,\Omega) \quad .
\end{equation}
Let us repeat the process $Y$ in $N$ independent experiments to obtain a data set $y = (y_1, \cdots, y_N)$. $k=(k_1,\cdots,k_W)$ is the histogram of the events recorded in $y$, i.e. $k_i$ is the number of times $z_i$ appears in $y$. Note that  $\sum_{i=1}^W k_i=N$. As a consequence of independent sampling, the probability to sample exactly $k$ is, 
\begin{equation}
	P(k|\theta,\Omega) = {N\choose{k}}\prod_{i=1}^W p(z_i|\theta,\Omega)^{k_i}\quad ,
\end{equation}
where ${N\choose{k}}=N!/\prod_{i=1}^W k_i!$ is the multinomial factor. 
Bayes' formula allows us to get an estimator for the parameters $\theta$,
\begin{equation}
	P(\theta|k,\Omega) = P(k|\theta,\Omega)\frac{P(\theta|\Omega)}{P(k|\Omega)}\,.
\label{bayes}
\end{equation}
Obviously, $P(k|\Omega) = \int d\theta\ P(k|\theta,\Omega)P(\theta|\Omega)$ does not depend on $\theta$. Without further available information we must assume that the parameters $\theta$ are uniformly distributed between their upper and lower limits. As a consequence, $P(\theta|\Omega)$ also does not depend on $\theta$ within the limits of the parameter range and can be treated as a constant\footnote{Unfortunately, what works for parameters in $\theta$ such as $\lambda$ does not work for parameters such as $x_{\rm min}$ and $x_{\rm max}$. For those variables it turns out that $P(\theta|\Omega)$ can not be assumed to be constant between upper and lower bounds of the respective parameter values. Bayesian estimators for $x_{\rm min}$ and $x_{\rm max}$ require to explicitly consider a non-trivial function $P(\theta|\Omega)$. Though in principle feasible, we ignore the possibility of deriving Bayesian estimates for $x_{\rm min}$ and $x_{\rm max}$ in this paper.
}.
From Eq. (\ref{bayes}) it follows that the value $\theta^*$ that maximizes $P(\theta|k,\Omega)$ also maximizes $P(k|\theta,\Omega)$. The most likely parameter values $\theta^*=(\theta_1,\cdots,\theta_R)^*$ are now found by maximizing the log-likelihood, 
\begin{equation}
\begin{array}{lcl}
	0&=&\frac{\partial}{\partial \theta_r}\frac{1}{N}\log P(\theta|k,\Omega)\\
	&=& \sum_{i=1}^W f_i\frac{\partial}{\partial \theta_r}\log 	 	p(z_i|\theta,\Omega)\\
	&=& -\frac{\partial}{\partial \theta_r} H_{\rm cross} 		(f||p(z|\theta,\Omega))\quad , 
\end{array}
\label{bayes3}
\end{equation}
for all parameters $r=1,\cdots,R$. Here $H_{\rm cross} (f||p(z|\theta,\Omega))\equiv-\sum_{i=1}^W f_i\log p(z_i|\lambda,\Omega)$, 
is the so-called {\em cross-entropy}. In other words, ML-estimates maximize the cross-entropy with respect to the parameters $\theta_r$.

\subsection{The ${\rm ML}^*$-algorithm for power-laws}

To apply Eq. (\ref{bayes3}) for ML-estimates of power-law exponents, one specifies the finite sample space $\Omega=\{z_1,z_2,\cdots,z_W\}$, and the family of probability density functions is, 
\begin{equation}
	p(x|\lambda,\Omega)=\frac{x^{-\lambda}}{Z_\lambda(\Omega)} \quad ,
\end{equation}
with $x\in\Omega$. Note that the set of parameters $\theta$ defined above now only contains $\lambda$, or $\theta=\{\lambda\}$.
The normalization constant is $Z_\lambda(\Omega)=\sum_{x\in\Omega} x^{-\lambda}$. The derivative with respect to $\lambda$ of the cross-entropy, 
$H_{\rm cross}(f||p(z|\theta,\Omega))=\lambda\sum_{i=1}^W f_i\log z_i+\log Z_\lambda(\Omega)$,
has to be computed, and setting $dH_{\rm cross}/d\lambda=0$ yields
\begin{equation}
	\sum_{i=1}^W f_i\log z_i=
	\left(\sum\limits_{i=1}^W z_i^{-\lambda}\right)^{-1}
	\sum\limits_{i=1}^W z_i^{-\lambda}\log z_i\quad .
\label{bayes4}
\end{equation}
The solution to this implicit equation, $\lambda=\lambda^*$, can not be written in closed form but can be easily solved numerically. See \cite{rplfit} for the corresponding algorithm and \app B for a tutorial.

%%%%%%%%%%%%%%%%%%%%%%%%%%%%%%%%%%
\subsection{How to determine $\lambda^*$}
%%%%%%%%%%%%%%%%%%%%%%%%%%%%%%%%%%
One possibility to find the solution $\lambda=\lambda^*$ from the implicit equation Eq. (\ref{bayes4}), is to iteratively refine approximate solutions. For this, select $M+1$ values $\lambda$ from the interval $[\lambda_{\min},\lambda_{\max}]$, where $M$ is a finite fixed number, say  $M=100$. Those values may be chosen to be given by the expression
\begin{equation}
	\lambda_r(m)=\underline{\lambda}_r+
	\frac{m}{M} \left(\overline{\lambda}_r-\underline{\lambda}_r \right)\,,
\end{equation}
for $m=0,\cdots,M$. The parameters $\underline{\lambda}_r$ and $\overline{\lambda}_r$ are defined in the following way: First define $\underline{\lambda}_1=\lambda_{\min}$, and $\overline{\lambda}_1=\lambda_{\max}$, where $\lambda_{\max}$ and $\lambda_{\min}$ are parameters of the algorithm. Then define $\delta\lambda_1=\Delta\lambda/M$ with $\Delta\lambda=\lambda_{\max}-\lambda_{\min}$. If $\lambda_1(m^*_1)$ is the optimal solution of Eq. (\ref{bayes4}) for some $m^*_1$, then we can choose $\underline{\lambda}_2=\lambda_1(m^*_1)-\delta\lambda_1$, and $\overline{\lambda}_2=\lambda_1(m^*_1)+\delta\lambda_1$ and $\delta\lambda_2=2\delta\lambda_1/M$. One then continues by iterating $r$ times until $\delta\lambda_r<\ep$, where $\ep$ is the desired accuracy of the estimate of $\lambda^*$. As a consequence, the value $m^*_r$, for which $|\lambda^*-\lambda_r(m^*_r)|<\ep$ holds, optimally estimates $\lambda^*$ in the $r$'th iteration with an error smaller than $\ep$. Note that $\ep$ is the error of the ${\rm ML}^*$-estimator with respect to the exact value of the predictor $\lambda^*$, and is not the error of $\lambda^*$ with respect to the (typically unknown) value of the exponent $\lambda$ of the sampling distribution.

\begin{figure}[t]
\centering
	\includegraphics[width=0.95\columnwidth]{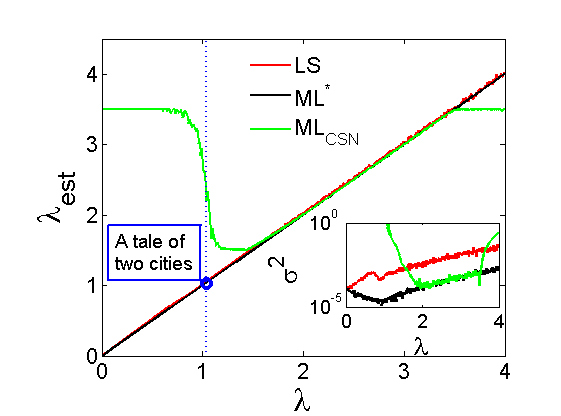}\\
\caption{Comparison of the three power-law exponent estimators, LS, ${\rm ML}_{\rm CSN}$, and ${\rm ML}^*$. For $400$ values of $\lambda$ in the range between 0 and 4, we sample $N=10,000$ events from $\Omega=\{1,\cdots,1,000\}$, from a power-law probability distribution $p(x|\lambda,\Omega)\propto x^{-\lambda}$. The estimated exponents $\lambda_{\rm est}$ for the estimators ${\rm LS}$ (red), the ${\rm ML}_{\rm CSN}$ (green, $\lambda_{\rm est}=\hat\lambda$), and the new ${\rm ML}^*$ (black, $\lambda_{\rm est}=\lambda^*$), are plotted against the true value of the exponent $\lambda$ of the probability distribution samples are drawn from. Clearly, below $\lambda\sim1.5$ the ${\rm ML}_{\rm CSN}$ estimator no longer works reliably. ${\rm ML}_{\rm CSN}$ and ${\rm ML}^*$ work equally well in a range of $1.5<\lambda<3.5$. Outside this range ${\rm ML}^*$ performs consistently better than the other methods. The inset shows the mean-square error $\sigma^2$ of the estimated exponents. The LS-estimator has a much higher $\sigma^2$ over the entire region, than the ${\rm ML}^*$-estimator. The blue dot represents the ${\rm ML}^*$ estimate for the Zipf exponent of C. Dickens' ``A tale of two cities''. Clearly, this exponent could never reliably be obtained  from the rank ordered distribution using ${\rm ML}_{\rm CSN}$, whereas ${\rm ML}^*$ works fine even for values of $\lambda\sim0$.	\label{fig:comparison}
}
\end{figure}

Controlling the fit region over which the power-law should be obtained therefore becomes a matter of restricting the sample space to a convenient $\Omega'\subset\Omega$. This can be used for dynamically controlling low-frequency cut-offs. These cut-offs are set to exclude states for which,
\begin{equation}
	p(z_i|\lambda,\Omega)N<k_{\min}\quad ,
\end{equation}
where $k_{\min}$ is the minimal number of times that any state-type $i$ is  represented in the data set. This means that we re-estimate $\lambda$ on $\Omega'\subset\Omega$ with 
\begin{equation}
	\Omega'=\{z_i\in\Omega|p(z_i|\lambda,\Omega)N \geq k_{\min}\}\quad .
\end{equation}
We see in Eq. (\ref{bayes4}) that iteratively adapting $\Omega$ to subsets $\Omega'$, and then re-evaluating $\lambda$, requires to solve,
\begin{equation}
		\sum\limits_{i\in I(\Omega')} f'_i\log\left(z_i\right)=
	\left(\sum\limits_{i\in I(\Omega')} z_i^{-\lambda}\right)^{-1}
	\sum\limits_{i\in I(\Omega')} z_i^{-\lambda}\log z_i\quad ,
\label{bayes5}
\end{equation}
where $N'=\sum_{i\in I(\Omega')}k_i$ is the restricted sample-size and $f'_i=k_i/N'$ are the relative frequencies re-normalized for $\Omega'$. $I(\Omega')=\{i|z_i\in\Omega'\}$ is the index-set of $\Omega'$.

Iterating this procedure either leads to a fixed point or to a limit cycle between two low-frequency cut-offs with two slightly different estimates for $\lambda^*$. These two possibilities need to be considered in order to implement an efficient stopping criterion for the iterative search of the desired low-frequency cut-off in the data. The algorithm therefore consists of two nested iterations. The ``outer iteration'' searches for the low-frequency cut-off, the ``inner iteration'' solves the implicit equation for the power-law exponent. The matlab code for the algorithm is found in \cite{rplfit}, see \app B for a tutorial.

%%%%%%%%%%%%%%%%%%%%%%%%%%%%%%%%%%
%%%%%%%%%%%%%%%%%%%%%%%%%%%%%%%%%%
\section{Testing the new estimator with numerical experiments and known data sets}
%%%%%%%%%%%%%%%%%%%%%%%%%%%%%%%%%%
%%%%%%%%%%%%%%%%%%%%%%%%%%%%%%%%%%

To test the proposed algorithm implementing the estimator ${\rm ML}^*$, we first perform numerical experiments and then test its performance on a number of well known data sets. 

\subsection{Testing with numerical experiments}

For 400 different values of $\lambda$, ranging from $0$ to $4$, we sample $N=10,000$ data points $x\in\Omega=\{1,\cdots,W\}$, with $W=1000$ states, with probabilities $p(x|\lambda,\Omega)\propto x^{-\lambda}$. We fit the data in three ways, using (i) least square fits (LS), (ii) the CSN algorithm ${\rm ML}_{\rm CSN}$  providing estimates $\hat\lambda$, and (iii) the implicit ${\rm ML}^*$ method providing estimates $\lambda^*$. In Fig. \ref{fig:comparison} we show these estimates for the power exponents, as a function of the true {\em values} of $\lambda$. The ${\rm LS}$, ${\rm ML}_{\rm CSN}$, ${\rm ML}^*$ estimators are  shown as the red, green, and black curves respectively. Obviously ${\rm ML}^*$ and ${\rm ML}_{\rm CSN}$ work equally well for power-law exponents $\lambda$ with values $1.5<\lambda<3.5$. In this range the three approaches coincide. However, note that  in the same region the mean square error\footnote{The mean square error is defined as  
$\sigma^2(\lambda)=N_{\rm rep}^{-1}\sum_{m=1}^{N_{\rm rep}} (\lambda_{\rm est}(m) -  \lambda)^2$, where $N_{\rm rep}$ is the number of repetitions, i.e. the number of data-sets we sampled from the $p(x|\lambda,\Omega)$, $x=1,\dots,W$. $\lambda_{\rm est}(m)$ is the value estimated for $\lambda$ from the $m$th data set. Depending on the estimator $\lambda_{\rm est}$ corresponds to $\hat\lambda$ (${\rm ML}_{\rm CSN}$), $\lambda^*$, (${\rm ML}^*$), or the LS estimator. We used $W=1000$ and $N_{\rm rep}=25$ for any given $\lambda$. 
} 
$\sigma^2$ for the LS method is much larger than for  ${\rm ML}^*$ and ${\rm ML}_{\rm CSN}$. Outside this range the assumptions and approximations used for ${\rm ML}_{\rm CSN}$ start to lose their validity and both ${\rm LS}$ and ${\rm ML}^*$ estimates outperform the ${\rm ML}_{\rm CSN}$ estimates. The inset also shows that ${\rm ML}^*$ consistently estimates $\lambda$ much better than the ${\rm LS}$ estimator (two orders of magnitude better in terms of $\sigma^2$) for the entire range of $\lambda$. 
The blue dot in Fig. \ref{fig:comparison} represents the ${\rm ML}^*$ estimate for the Zipf exponent of C. Dickens' `A tale of two cities'. Clearly, this small exponent could never be obtained by ${\rm ML}_{\rm CSN}$, see also Tab. \ref{table1}. 

\begin{table}[t]
\begin{center}
\begin{tabular}{l c c c c c c}
  & exp. & ${\rm CSN}_1$ & ${\rm CSN}_2$ & ${\rm ML}^*$ & ${\rm KS}_{\rm CSN}$ & ${\rm KS}^*$ \\
 \hline
blackouts & $\lambda$ & 2.3 & 2.27 & 2.25 & 0.061 & 0.031 \\
surnames & $\alpha$  & 2.5 & 2.49 & 2.66 & 0.041 & 0.019 \\
int. wars &  $\lambda$ & 1.7 & 1.73 & 1.83 & 0.078 & 0.076 \\
city pop. &   $\lambda$ & 2.37 & 2.36 & 2.31 & 0.019 & 0.016 \\
quake int. & $\lambda$ & 1.64 & 1.64 & 1.88 & 0.092 & 0.085 \\
relig. fol. & $\lambda$   & 1.8 & 1.79 & 1.61 & 0.091 & 0.095 \\
citations & $\lambda$   & 3.16 & 3.16 & 3.10 & 0.010 & 0.018 \\
words & $\alpha$       & 1.95 & 1.95 & 1.99 & 0.009 & 0.015 \\
wealth & $\lambda$      & 2.3 & 2.34 & 2.30 & 0.063 & 0.066 \\
papers & $\lambda$      & 4.3 & 4.32 & 3.89 & 0.079 & 0.082 \\
sol. flares & $\lambda$  & 1.79 & 1.79 & 1.81 & 0.009 & 0.021 \\
terr. attacks & $\lambda$  & 2.4 & 2.37 & 2.36 & 0.018 & 0.017 \\
websites & $\lambda$        & 2.336 & 2.12 & 1.72 & 0.025 & 0.056 \\
forest fires & $\lambda$      & 2.2 & 2.16 & 2.46 & 0.036 & 0.034 \\
Dickens novel & $\lambda$ & - & - & 1.04 & - & 0.017 \\
\end{tabular}
\end{center}
\caption{Comparison of the estimators ${\rm ML}^*$ and ${\rm ML}_{\rm CSN}$ on empirical data sets that were used in \cite{plreview}. These include the frequency of surnames, intensity of wars, populations of cities, earthquake intensity, numbers of religious followers, citations of scientific papers, counts of words, wealth of the Forbes 500 firms, numbers of papers authored, solar flare intensity, terrorist attack severity, numbers of links to websites, and forest fire sizes. We added the word frequencies in the novel ``A tale of two cities" (C. Dickens). The second column states if $\alpha$ or $\lambda$ were estimated. The exponents reported in \cite{plreview} are found in column ${\rm CSN}_1$, those reproduced by us applying their algorithm to data \cite{plreview,aux1,aux2,aux3,aux4} is shown in column ${\rm CSN}_2$. The latter correspond well with the new ${\rm ML}^*$ algorithm. For values $\lambda<1.5$, ${\rm CSN}$ can not be used. We list the corresponding values for Kolmogorov-Smirnov test for the two estimators, ${\rm KS}_{\rm CSN}$ and ${\rm KS}^*$.  
\label{table1}
}
\end{table}

\subsection{Testing with empirical data sets}

We finally compare the new estimator ${\rm ML}^*$ on several empirical data sets that were used for demonstration in \cite{plreview}. In Tab. \ref{table1} we collect the results. The second column states if $\lambda$ or $\alpha$ were estimated.
Column ${\rm CSN}_1$ presents the value of the estimator ${\rm ML}_{\rm CSN}$ as presented in \cite{plreview}. 
Column ${\rm CSN}_2$ contains the values of the same estimator using the data from \cite{plreview} and using the algorithm provided by \cite{Clcode}\footnote{The reason for the differences might be that some of the data has been updated since the publication.}. The results for the ${\rm ML}^*$ estimator agrees well with those of ${\rm ML}_{\rm CSN}$ in the range where the latter works well. To demonstrate how ${\rm ML}^*$ works perfectly outside of the comfort zone of ${\rm ML}_{\rm CSN}$ (for $\lambda<1.5$), we add the result of the rank distribution of word counts in the novel ``A tale of two cities" (Charles Dickens, 1859), which shows an exponent of $\lambda\sim1.035$. This exponent can be fitted directly from the data using the proposed ${\rm ML}^*$ algorithm, while ${\rm ML}_{\rm CSN}$ can not access this range, at least not without the detour of first producing a histogram from the data and then fitting the tail of the frequency distribution. The values for the corresponding Kolmogorov-Smirnov tests (see e.g. \cite{plreview}) for the two estimates, ${\rm KS}_{\rm CSN}$ and ${\rm KS}^*$, are similar for most cases. 

\section{Conclusions}

We discuss the generic problem of estimating power-law exponents from data sets. We list a series of questions that must be clarified before estimates can be performed. 
We present these questions in form of a decision tree that shows how the answers to those questions lead to different strategies for estimating power-law exponents. 

To follow this decision tree can be seen as a recipe for fitting power exponents from empirical data. The corresponding algorithms were presented and can be downloaded as matlab code. The two algorithms we provide are based on a very general ML estimator that maximizes an appropriately defined cross entropy. The method can be seen as a straight forward generalization of the idea developed in \cite{plreview}. The two estimators (one for binned histograms and ${\rm ML}^*$ for raw data sets) allow us to estimate power-law exponents in a much wider range than was previously possible. In particular, exponents lower than $\lambda<1.5$ can now be reliably obtained. 

\section*{Acknowledgments}
\noindent This work was supported in part by the Austrian Science Foundation FWF under grant P29252. 
B.L. is grateful for the support by the China Scholarship Council, file-number 201306230096.
 
%\nolinenumbers

% Either type in your references using
% \begin{thebibliography}{}
% \bibitem{}
% Text
% \end{thebibliography}
%
% or
%
% Compile your BiBTeX database using our plos2015.bst
% style file and paste the contents of your .bbl file
% here.
% 
%%%%%%%%%%%%%%%%%%%%%%%%%%%%%%%%%%%%%%%%%%%%%%%%%%%%%%%%  

%%%%%%%%%%%%%%%%%%%%%%%%%%%%%%%%%%%%%%%%%%%%%%%%%%%%%%%%  

\newpage
%\appendix*

\section*{APPENDIX A: Sampling from continuous sample spaces}\label{contspace}

If events $x$ are drawn from a continuous sample space $\Omega=[x_{\rm min},\,x_{\rm max}]$,  for instance the magnitude of earthquakes, then the `natural order' of possible events is simply given by the magnitude $x$ of the observation. Events $x$ are drawn from a continuous power-law distribution $p(z|\lambda,\Omega)=x^{-\lambda}/Z$, with $Z=Z_\lambda([x_{\rm min},\,x_{\rm max}])$ (compare Eq. (\ref{norm}) first line).

To work with well defined probabilities we have to bin the data first. Probabilities to observe events within a particular bin depend on the margins of the $W$ bins $b=(b_0,b_1,\cdots,b_W)$, with $b_0=x_{\rm min}$ and $b_W=x_{\rm max}$. The histogram $k=(k_1,\cdots,k_W)$ counts the number $k_i$ of events $x$ falling into the bin $b_i>x\geq b_{i-1}$, and the probability of observing $x$ in the $i$'th bin is given by
\begin{equation}
	p(i|\lambda,x)=\frac{b_{i}^{1-\lambda}-b_{i-1}^{1-\lambda}}{x_{\rm max}^{1-\lambda}-x_{\rm min}^{1-\lambda}}\quad .
	\label{appa1}
\end{equation}
Binning events sampled from a continuous distribution may have practical reasons. For instance data may be collected from measurements with different physical resolution levels, so that binning should be performed at the
lowest resolution of data points included in the collection of samples. We will not discuss the ML estimator for binned data in detail but only remark that for given bin margins $b$ it is sufficient to insert $p(i|\lambda,x)$ of Eq (\ref{appa1}) into Eq. (\ref{bayes3}) with $\theta=\{\lambda\}$, to derive the appropriate ML condition for binned data. An algorithm for binned data \verb|r_plhistfit|, where we assume the bin margins $b_i$ to be given, is found in \cite{rplfit}.

We point out that if margins for binning have not been specified prior to the experiments, then specifying the optimal margins for binning the data becomes a parameter estimation problem in itself, i.e. the optimal margins $b_i$ have to be estimated from the data as well. One major source of uncertainty in the estimates of $\lambda$ from binned data is related to the uncertainty in choosing the upper and lower bounds $x_{\rm min}$ and $x_{\rm max}$ of the data, i.e. specifying the bounds of the underlying continuous sample space.

Binning becomes irrelevant for clean continuous data for the following reason. Suppose we fix the sample space $[x_{\rm min},\ x_{\rm max}]$ and cut this domain into $M$ bins of width $\Delta=(x_{\rm max}-x_{\rm min})/M$. Since the data $x=\{x_1,\cdots,x_N\}$ is drawn from a continuous sample space, 
the chance for two observations $x_m$ and $x_n$ to be exactly equal becomes zero for $m\neq n$, if $M$ has been chosen sufficiently large. Then each bin almost certainly contains either one sample $x_n$ or none. The probability of observing $x$ then is asymptotically (as $\Delta$ approaches zero) given by 
\begin{equation}
	P(x|\lambda)=\Delta^N\prod\limits_{n=1}^N \left(\frac{x_n^{-\lambda}}{Z_\lambda(x_{\min},x_{\max})}\right) \quad .
\end{equation}
The parameter estimation problem of finding the optimal $\lambda$ is equivalent to maximizing $P(x|\lambda)$ (or equivalently $\log P(x|\lambda)$) with respect to $\lambda$. In this maximization problem $\Delta$ becomes irrelevant and only the choice of $x_{\min}$ and $x_{\max}$ and the data $x$ remains relevant for the estimate. As a consequence, one obtains an equation 
\begin{equation}
	\sum_{i=1}^W f_i\log z_i=\frac{d}{d\lambda}\log Z_\lambda\,,
\label{bayesX}
\end{equation}
for the ML estimate of the exponent $\lambda$ over continuous sample spaces. Equation (\ref{bayes4}) and Eq. (\ref{bayesX}) differ only in $Z_\lambda$. In Eq. (\ref{bayes4}) the normalization constant of discrete samples spaces gets used while in Eq. (\ref{bayesX}) $Z_\lambda$ is the normalization constant for a continuous sample space. Switching between continuous and discrete sample spaces therefore is simply a matter of choosing the one or the other normalization constant in the algorithm.

Whether data should be assumed to be sampled from continuous or discrete sample spaces is not always totally clear. Many measurements have an intrinsic resolution and implicitly bin the data. For instance if real numbers sampled in an experiment 
are given only with a three digit precision, such as $x_n=0.123$ and we know that $0.001=x_{\rm min}$ and $x_{\rm max}=5$ then we better treat the data as discrete data on $\Omega_d=\{0.001,0.002,\cdots,4.998, 4.999,5\}$ if we have sufficiently many samples for the histogram over $\Omega_d$ not to be flat.
A primitive test to see whether one should regard data as sampled from a continuous sample space or not is to make a histogram over the unique values of the recorded data. If each distinct value appears only once in the data (i.e. if the histogram over the unique data-points is flat) then one should treat the sample-space as continuous.

While for the discrete case we need not estimate $x_{\rm min}$ and $x_{\rm max}$ this remains necessary for the continuous case. The method of cutting the $[x_{\rm min},x_{\rm max}]$ into segments of length $\Delta$ and then taking $\Delta$ to zero explains why typically tha {\em primitive} estimates, $x_{\rm min}=\min\{x_n|n=1,\cdots,N\}$ and $x_{\rm max}=\max\{x_n|n=1,\cdots,N\}$, provides fairly good results. Alternatively, strategies such as suggested in \cite{plreview} could be used to optimize the choices for $x_{\rm min}$ and $x_{\rm max}$. However, this procedure can not be directly derived from Bayesian arguments. Neither will we discuss this approach in this paper nor implement such an option in \verb|r_plfit|.

However, Bayesian estimates of $x_{\rm min}$ and $x_{\rm max}$ exist. Although we will not discuss those estimators in detail here we will eventually implement them in \verb|r_plfit| to replace the primitive estimates. The idea of constructing such estimators is the following. For instance, one asks how likely can the maximal value $\max(x)$ of the sampled data $x=(x_1,\cdots,x_N)$ be found to be larger than some value $y$. By deriving $P(\max(x)>y|\lambda,[x_{\rm min},x_{\rm max}])$ and $P(\min(x)<y|\lambda,[x_{\rm min},x_{\rm max}])$, as a consequence, it becomes possible to derive Bayesian estimators for $x_{\rm min}$ and $x_{\rm max}$. 

%%%%%%%%%%%%%%%%%%%%%%%%%%%%%%%%%%%%%%%%%
\section*{APPENDIX B: Using r\_plfit}
%%%%%%%%%%%%%%%%%%%%%%%%%%%%%%%%%%%%%%%%%
The matlab function
\begin{center}
\verb|function out = r_plfit(data,varargin)|
\end{center}
implements the algorithm discussed in the main paper. The function returns a struct \verb|out| that contains information about the data, the data range, but most and for all \verb|out.exponent| returns the estimated exponent of the power-law. Whether the exponent \verb|out.exponent| is the exponent $\lambda$ of the sample distribution or the exponent $\alpha$ of the frequency distribution of the data depends on how \verb|function out = r_plfit(data,varargin)| gets used as explained below. In the code the sample space $\Omega$ is equivalent to a vector $z=[z_1,\cdots,z_W]$ containing $W$ distinct event magnitudes $z_i$, $i=1,\cdots,W$. \\

The variable \verb|data| can be used to import data while a variable number of arguments can be set by \verb|varargin| to tell the algorithm which type of data it should handle and to control the range of the data. By default the only argument that has to be set is \verb|data|. \verb|r_plfit| filters data from data points \verb|data<=0, NaN, Inf|. The data passed on to \verb|data| can be
\begin{itemize}
\item a vector of observations \verb|data| $\equiv x=[x_1,\cdots,x_N]$ (default) 
\item a histogram \verb|data| $\equiv k=[k_1,\cdots,k_W]$ of recorded event types $i=1,\cdots,W$ 
\end{itemize}
\verb|out = r_plfit(data,varargin)| can be used in three basic modes
\begin{itemize}
\item \verb|out = r_plfit(x)| returns the estimated exponent $\lambda$ of the {\bf probability distribution} given the observation $x$ (default) 
\item \verb|out = r_plfit(k,'hist')| returns the estimated exponent $\lambda$ of the {\bf probability distribution} given the histogram of observations $k$
\item \verb|out = r_plfit(k)| returns the estimated exponent $\alpha$ of the {\bf frequency distribution} given the histogram of observations $k$
\end{itemize}
The third mode \verb|out = r_plfit(k)| is in fact identical to the first mode \verb|out = r_plfit(x)|, only that passing a histogram as sample data to the algorithm is identical to asking how many of the $W$ states $i$ have been observed $n$ times. But this is exactly the frequency distribution of the process, which possesses a tail with exponent $\alpha=1+1/\lambda$. Depending on the mode \verb|r_plfit| returns the exponent $\lambda$ or $\alpha$ in \verb|out.exponent| \\

{\bf Fitting with observations $x$}: If we run \verb|out = r_plfit(x)| without further options \verb|r_plfit| assumes by default that the data $x$ consists of natural numbers, and that the process samples have been sampled from the sample space $\Omega=\{\min(x),\min(x)+1,\cdots,\max(x)-1,\max)\}$, i.e. $\min(x)\leq z_i=i\leq\max(x)$. If this is not the case one can either specify the data range using all $W$ {\em unique} values $z=[z_1,\cdots,z_W]$ occurring in the data $x$ by using the option \verb|out = r_plfit(x,'urange')|. In order to define a fit range maximal and minimal data values taken into account can be set by \verb|out = r_plfit(x,'urange','rangemin',minval, ...| \verb|... 'rangemax',maxval)| such that \verb|r_plfit| only takes into account data in the range \verb|minval| $\leq z\leq$ \verb|maxval|. To control the data range individually use \verb|out = r_plfit(x,'range',z)|. If the data has been sampled from a continuous sample space, and the histogram over the unique data is flat, i.e. each value in the data only appears once (more or less), then one can tell \verb|r_plfit| that the data is sampled from a continuous sample space by setting the option \verb|'cdat'|, i.e. by running \verb|out = r_plfit(x,'cdat', ...)|. This option tells the algorithm to use the normalization constant for continuous sample spaces and estimates $x_{\rm min}=\min(x)$ and $x_{\rm max}=\max(x)$. Moreover, \verb|'cdat'| implicitly sets the \verb|'urange'| and the \verb|'nolf'| option. \verb|'nolf'| (see below) switches off the search of the algorithm for an optimal low frequency cut-off. \\

{\bf Fitting with histograms $k$}: Using histograms $k$ as input works in exactly the same way as for fitting $x$ if we want to estimate the exponent $\alpha$ of the frequency distribution and use \verb|r_plfit| in the \verb|out = r_plfit(k)| mode. If we use \verb|r_plfit| in the \verb|out = r_plfit(k,'hist')| mode, the algorithm assumes by default that the sample space $z$ is given by $z=[1,2,\cdots,W]$. The option \verb|'urange'| has no effect in this mode and gets ignored if set. Otherwise one can again use the \verb|'range'| property to set the event magnitudes $z$ (the sample space) using \verb|out = r_plfit(k,'hist','range',z)|. The \verb|'minrange'| and \verb|'maxrange'| options work in exactly the same way as before.\\

{\bf Dynamic low frequency cut-off}: By default \verb|r_plfit(data)| runs an iterative search for an optimal low frequency cut-off that is set at a range value $z_i$ such that the expected number of samples for $z_i$ equals the variable $N_{\rm min}$ (default value $1$, reset using option \verb|'Nmin'|). This means the algorithm performs a low frequency cut-off for observations $x$. If however \verb|maxval| is smaller than the predicted cut-off then the low frequency cut-off has no effect. One should note that in the mode \verb|out = r_plfit(k)| the low frequency cut-off mechanism effectively acts as a high frequency cut-off with respect to the data $x$. One can switch this mechanism off by setting the option \verb|'nolf'| (no low frequency cut-off).\\

The \verb|'plot'| option, \verb|out = r_plfit(data,...| \verb|...,'plot')|, can be used for visualization. \verb|r_plfit| plots the fit over the data in double logarithmic coordinates (loglog plot). Using the option \verb|'figure'| behaves like \verb|'plot'| but explicitly opens a new figure. \verb|'exp_min'| can be used to specify the minimal search value for the exponents (default is $0$) and \verb|'exp_max'| to set the maximal search value (default is $5$). \verb|'eps'| can be used to set the precision of the implicit algorithm (default $1e-5$). Several other options exist to control the performance of the algorithm, which all can be listed by using \verb|r_plfit('help')| in the command line, which prints a brief manual on the usage of \verb|r_plfit| and available options.

%%%%%%%%%%%%%%%%%%%%%%%%%%%%%%%%%%%%%%%%%
\section{ Using r\_plhistfit}
%%%%%%%%%%%%%%%%%%%%%%%%%%%%%%%%%%%%%%%%%

If one works with binned data, e.g. histogram data counting the number of events falling into exponentially scaled bins (log-binning), then \verb|r_plhistfit| needs to be used instead of \verb|r_plfit|. The function \verb|function out = r_plhistfit(data,varargin)| like \verb|r_plfit|, by default, uses only data as input and other variables can be set optionally. \verb|data| is always a histogram $k$ that is a vector $k=[k_1,\cdots,k_W]$. Bins can be specified by giving bin margins $b=[b_0,b_1,\cdots,b_W]$ such thatevents counted in $k_i$ had a magnitude $x$ such that $b_{i-1}\leq x<b_{i}$. Usage, \verb|r_plhistfit(k,'margins',b)|. By default \verb|r_plhistfit| assumes that $b_i=i+1/2$. Other options work similar to the ones available for \verb|r_plfit| and can be reviewed by typing \verb|r_plhistfit('help')| in the matlab command line.

\end{document}